\documentclass{article}

\usepackage{fullpage}

\usepackage{amssymb}
\usepackage{amsmath,amsfonts,bm}
\newcommand{\be}{\begin{equation}} \newcommand{\ee}{\end{equation}}
\newcommand{\ba}{\begin{eqnarray}} \newcommand{\ea}{\end{eqnarray}}
\newcommand{\ban}{\begin{eqnarray*}}
\newcommand{\ean}{\end{eqnarray*}}
\newcommand{\barr}{\be\left\{\begin{array}}
\newcommand{\earr}{\end{array}\right.\ee}

\def\pmb#1{\setbox0=\hbox{$#1$}\kern-.025em\copy0\kern-\wd0
\kern-0.05em\copy0\kern-\wd0\kern-.025em\raise.0233em\box0}
   
\newcommand{\half}{\mbox{$\frac{1}{2}$}}

\title{On The Right Side of Einstein's Equation}

\date{}

\author{Edward A. Spiegel\footnote{Visiting Scholar, 
New York University, New York, NY 10003 USA}
\vspace{1mm}\\Department of Astronomy\\Columbia University\\
New York, NY 10027 USA\,}

\begin{document}
\maketitle

\centerline {Based on two lectures in the
49th Winter School in Theoretical Physics Karp49 }
\centerline{(Cosmology and non-equilibrium statistical mechanics)
Ladek Zdr\'oj, Poland, February 10-16.} 

\centerline {To appear in IJGMMP in 2014 with minor changes herein.}

\begin{abstract}
Recent developments in observational cosmology have led to 
attempts to make modifications on both sides of the Einstein equation 
to explain some of the puzzling new findings.  
What follows is an examination of the source of gravity that we usually find on the right hand side of Einstein's equation.   The outcome is a 
modified version of the stress-energy tensor that is the source of the gravitational field.
The derivation is based on the kinetic theory of a gas of identical particles with no internal structure. 

The presentation here is in two parts.  In {\bf Part I}, I describe the stress tensor that Xinzhong Chen and I have proposed for the matter tensor for a nonrelativistic gas
with input from Hongling Rao and Jean-Luc Thiffeault. 
Our derivation of the equations of fluid dynamics is based on kinetic theory without recourse to the standard Chapman-Enskog approximation.
The nonrelativistic treatment reveals the underlying physics clearly and it facilitates comparison with experimental results on acoustic propagation.  Further, it provides a setting for a Newtonian cosmology, which remains a topic of some interest.  I regret though that there is room only for the relativistic version of the treatment 
of cosmology.

In {\bf Part II}, I present the analogous derivation of our form for the stress-energy tensor in the relativistic case.  Then I exhibit its application to the usual isotropic cosmological model.  The result of that, in addition to the Friedmann solution, is a second solution that arises from terms discarded in the usual Chapman-Enskog approximation.  The new solution is a temporal analogue of a spatial shock wave.

Just as the usual shock waves make transitions in properties within
a mean free path, the new solution can change its properties appreciably in a mean flight time.   Whereas the Friedmann solution is not dissipative, the new solution produces entropy at a rate that may be of cosmological interest.  For the calculation of cosmic entropy production I use a formula derived in the ultrarelativistic limit in which particle masses are negligible.  A stability calculation to decide if and when the new solution may be realized has not been done since the microphysics is so uncertain.

Independently of the cosmological aspects, the fluid dynamical equations that we derive are causal, even for the heat equation
(or Fourier equation).
\end{abstract}
\medskip

keywords: kinetic theory; fluid dynamics; causality; cosmology.
\newpage

\section*{\centerline{\bf I. Nonrelativistic Theory}}

 \section{Definitions of the Basic Quantities}
 \subsection{The phase-space density}
 Since I have come to Poland to deliver these lectures I have developed a deeper understanding of the meaning of the word 
{\it soup}.  I can now call on this understanding when I say that
every astrophysical body is made up of a rich soup of different kinds of particles.  It is therefore with trepidation that I pretend that the simplified model of gases that I next adopt may capture 
the essential physics of the processes of interest 

I propose to discuss the dynamics
of a gas of identical particles, each of mass $m$.  These particles have no internal structure and, in {\bf Part I} only, obey Newtonian physics with infinite speed of light and zero Planck's
constant.  Particles may differ only in position, ${\bf x}$, and velocity, ${\bf v}$.  Each particle is represented by 
a point in a six-dimensional (Euclidean) phase space with coordinates  
${\bf x}$ and ${\bf v}$.  The expected number of particles in an infinitesimal volume element $d{\bf x} d{\bf v}$ is \begin{equation}
dN = f({\bf x, v}, t) \, d{\bf x} d{\bf v}
\end{equation}
where $f$ is the phase-space density
function that is central to our subject since the important macroscopic quantities arise as moments of $f$ over particle velocities \cite{uhl63}.
\subsection{Macroscopic quantities}
Within the infinitesimal spatial volume element, $d{\bf x}$,
the expected number of particles 
($=nd{\bf x}$) is
obtained by summing over all velocities.  The mass of this morsel of fluid, or fluid element, is found in this way and so we can obtain the
mass density of particles in space ($\rho = mn$) from \footnote{Integrals in velocity space herein are over all ${\bf v}$.}   \begin{equation}
\rho({\bf x}, t) = m\int f({\bf x}, {\bf v}, t)\, d{\bf v}\, . \label{rho}
\end{equation}
Similarly, the space density of the 
momentum is \begin{equation}
\rho {\bf u} = m\int {\bf v} f d{\bf v} . \label{mo}
\end{equation} 
The mean velocity of the particles in the fluid contained in 
$d{\bf x}$ is ${\bf u}({\bf x}, t)$ and the particle velocity with
respect to the local reference frame moving with the velocity 
${\bf u}$, the {\it peculiar velocity}, is \be 
{\pmb{\xi}} = {\bf v} - {\bf u}. \label{xi}
\ee
The tendency of the particles to
disperse is characterized by the temperature,
$T({\bf x}, {\bf v}, t),$  defined by \begin{equation}
\rho\,\Re\, T = \frac{1}{3}m \int {\pmb{\xi}}^2 f d{\bf v} \label{T}
 \end{equation}
where the gas constant is $\Re = k/m$ and $k$ is Boltzmann's
constant.  The fluid element must be so small that variations of physical quantities within it may neglected.
Two other moments of $f$, the pressure tensor and the heat flux,
will appear in the sequel:
\begin{equation}
{\mathbb P} =m\int {\bf c c} f d{\bf v}\, ,
\label{presst}
\end{equation}
 \begin{equation}
{\bf Q} ={\half}{m}\int {\bf c}^2 {\bf c} f d{\bf v}\, .
\label{heatf}
\end{equation}
\section{The kinetic equation}
Since $f$ is a density (in phase space) it is
governed by the continuity equation, 
\begin{equation}
\partial_t f + \partial_{i} \left(\dot x^i f \right) 
+ \frac{\partial}{\partial v^j}\left(\dot v^j f \right) = {\mathcal C}\left[f \right].
\end{equation}
Here $\partial_t := \partial/(\partial t)$ and 
$\partial_i :=\partial/\partial {x^i}$;
the $x^i$ and the $v^j$ are the components of ${\bf x}$
and of ${\bf v}$ with $i,j = 1,2,3$ and repeated indices
are summed over.    

The operator ${\mathcal C}$ represents the action of the medium
on individual particles passing through it and it contains the
essence of the kinetic model \cite{fer72}.  An equation of this form may be applied to a variety of physical problems such as the
passage of photons through a plasma or the diffusion of  neutrons
through a reactor.  Here we consider that the particles of interest travel through a medium made entirely of particles like itself.  In Boltzmann's model the influence of the ambient medium on a particle takes place only through its interaction with another  particle.  Boltzmann assumed that the members of each pair of interacting particles have not had a prior encounter so that the rate of encounters is proportional to the product of the the two relevant distribution functions.   

Without being explicit about the details of these interactions, we
let $j$ (not an index) be the rate at which particles not in the infinitesimal volume containing ${\bf v}$ are sent into this volume; 
$j$ may be thought of as a transition probability.  The rate
at which particles already in the infinitesimal volume are ejected 
is $\varkappa f$.  In Boltzmann's model $\varkappa$ is a linear functional of $f$.

We assume that the particles in question obey Hamiltonian
dynamics, so that \begin{equation}
 \partial_{i}\dot x^i  
+ \frac{\partial \dot v^j}{\partial v^j} = 0 \end{equation} 
(with $i,j=1,2,3$).  Hence the kinetic equation becomes \begin{equation}
\partial_t f + v^i \partial_i f + a^\ell \frac{\partial f}{\partial v^\ell}
= j - \varkappa f \, , \label{kineq}
\end{equation}
where $v^i=\dot x^i$ and $a^\ell=\dot v^\ell$ is an external force
per unit mass.
\section{The matching condition}
An $f = f_{0}$ for which the right hand side of (\ref{kineq})
vanishes is called a local equilibrium.  If the left hand side 
also vanishes, then we have a global (or true) equilibrium. 
 
In the case of Boltzmann's model, which is not spelled out here,  $f_{0}$ is found to be the Maxwell-Boltzmann distribution,
\begin{equation}
f_M=n(2\pi \Re T)^{-\frac{3}{2}}\exp(-\frac{{\pmb{\xi}}^2}{2\Re T}).
\label{f0}  
\end{equation}
In Boltzmann's model, ${\mathcal C}$ operates on ${\bf v}$ only so that, as far as it is concerned, ${\bf x}$ and $t$ are
parameters.  Therefore $f_{0}$ can depend on ${\bf x}$ and $t$
through its dependence on $\rho$, $T$ and ${\bf u}$. 

Since we express $f_{0}$ in terms of the current values of these thermodynamic quantities (including ${\bf u}$) and they in turn 
are expressed in terms of $f$ by (\ref{rho})-(\ref{T}), we impose the 
{\it matching conditions} 
\begin{equation}
\int \psi^A f d{\bf v} = \int \psi^A f_M d{\bf v}\, , \label{MC}
\end{equation}
with $A=0,...,4$ and 
\begin{equation}
\psi^0 = m, \ \ \ \ \psi^i =  mv^i, \ \ \ \ \psi^4 = \tfrac{1}{2}m{\bf v}^2.    \label{psi}
\end{equation}

The kinetic theory models that we contemplate here are assumed
to conserve the $\psi^A$.  That is, 
\begin{equation}
\int \psi^A  \left(j - \varkappa f_0\right) d{\bf v} = 0 . \label{int}
\end{equation}
\section{The equations of fluid dynamics}    
If we now multiply (\ref{kineq}) by $\psi^A$ and integrate over 
${\bf v}$, we obtain \begin{equation}
\partial_t \rho+\nabla\cdot (\rho {\bf u})=0 \label{ma}
\end{equation} \begin{equation}
 \partial_t {\bf u}+ {\bf u}\nabla {\bf u}- {\bf a}+\frac{1}{\rho}
\nabla\cdot{\mathbb P} =0 \label{mo}
\end{equation} \begin{equation}
 \partial_t T+{\bf u}\cdot\nabla T+\frac{2}{3\Re \rho}({\mathbb P}: 
\nabla {\bf u}+\nabla\cdot {\bf Q}) =0\, ,
\label{en}
\end{equation}
where the colon stands for a double dot product.
We assume that $f$ goes rapidly to zero at large speed and
that either ${\bf a}$ is given or that an equation for it is provided.

These continuum equations are a formal consequence of the kinetic equation and of our adoption of the the Maxwell-Boltzmann distribution as the local equilibrium subject to the matching condition.  Unfortunately, these
equations are not self-contained as we have as yet no prescription for determining $\mathbb P$ and ${\bf Q}$.  The question that lies open then is whether we can complete the system of equations (\ref{ma})-(\ref{en}) with closure representations for the undetermined quantities so that the resulting equations provide
a reasonably accurate representation of the behavior of the fluid.   
\section{Qualitative Considerations\protect\footnote{This discursive section is included to provide intuitive background.  It may be skipped without loss of comprehension of the sequel.}}
\subsection{A suggestive transformation}
The fluid equations (\ref{ma}-\ref{en})  remain to be completed by some prescription for finding the unknowns $\mathbb P$ and 
${\bf Q}$.   Evidently, we can obtain equations for these higher moments by taking higher moments of (\ref{kineq}).   But those additional equations will contain yet higher, but unprescribed, moments.  Hence the extension of this procedure, leads to an infinite sequence of moment equations that needs to be
terminated at some point.   The initial steps of the procedure are
called Grad's moment method \cite{HGrad49} but, effectively, the approach goes back some decades earlier to the work of Eddington and others in radiative transfer theory.  And it has been extended to higher moments in both contexts.
   
To see why extending this approach is not efficacious, we put the
kinetic equation into a form that is essentially a dynamical system.   To do this we introduce the new independent variables 
$\hat t, \hat{\bf x}$ and $\hat{\bf v}$ such that 
$t, {\bf x}$ and ${\bf v}$ depend on the new variables according to the equations  
\begin{equation}
\frac{dt}{d\hat t} = 1\ \ \ \ \frac{d{\bf x}}{d\hat t} = {\bf v}\ \ \ \ \frac{d{\bf v}}{d\hat t} = {\bf a}\, . \label{Ch} \end{equation}
Then as in the method of characteristics used to study partial differential equations \cite{john82}, we find that (\ref{kineq}) becomes \begin{equation}
\frac{d\hat f}{d\hat t} = {\mathcal C[\hat f]}\,  \label{EQ}\end{equation}
where \be
\hat f(\hat{\bf x}, \hat{\bf v},\hat t) = f({\bf x}, {\bf v}, t)   \ee
and ${\bf x}, {\bf v}, t$
depend on $\hat{\bf x}, \hat{\bf v},\hat t$.  We also add as a side condition the restriction to a characteristic curve on which 
$\hat{\bf x}$ and $\hat{\bf v}$ are constant.

On working with this formulation we find that the lower moments, 
$\rho, {\bf u}$ and $T$ are slow variables while the higher moments
are fast variables.  Hence including only a few higher moments does not make for great improvement in the derived macroscopic
equations while adding many moments calls for a lot of calculating to get results.  

I have introduced the transformation to characteristic coordinates  to bring out the similarity between kinetic theory and dynamical system theory.  But the additional complication caused by the restriction to a single characteristic curve would take us too far afield  if we were to develop the analogy further here.    While there may even be a center manifold theorem lurking behind this discussion, this qualitative account does not aspire to such heights.  All that is wanted from this digression is a suggestion as to why we retain only five moments of $f$ in the equations of fluid dynamics: the higher moments are fast variables and thus slavish.   
\subsection{Linear Theory}
Still in the pedagogical mode, let us make things simple by 
assuming that our system is homogeneous in space and 
seek solutions of the kinetic equation in the form \be
f({\bf v}, t) := 
f_0({\bf v})\left[1 + g({\bf v}, t\right]\, . \label{perturbed} \ee
Here $f_{0}$ satisfies (\ref{kineq}) and is a true (that is, global) equilibrium solution. 

On inserting (\ref{perturbed}) into (\ref{kineq}), we obtain an equation for $g$ which,  after some manipulation, takes the form\be
\frac{dg}{dt} = {\mathcal L}[g] + {\mathcal N}[g] , \label{u} \ee
where ${\mathcal L}[g]$ and ${\mathcal N}[g]$ are respectively linear and nonlinear terms in $g$ and the operators ${\mathcal L}$
and ${\mathcal N}$ depend on the properties of $f_{0}$ and on 
${\pmb{\nabla}}$. 

Much depends on the nature of the linearized collision operator of Boltzmann, so I first report some of its features while omitting
unneeded details \cite{foch70}.  The basic linear equation is \be
\frac{dg}{dt} = {\mathcal L}[g]\, . \label{lin} \ee
In Boltzmann's collision operator ${\mathcal C}$ there is no explicit dependence on time (though it may operate on functions of time) and therefore none in ${\mathcal L}$.  So we may seek separable solutions of the form \be 
g(t, {\bf v}) = \exp({\lambda t})\,
{\pmb{\Upsilon}}({\bf v}) \label{sep} \ee
where $\lambda$ is an eigenvalue of \be
{\mathcal L} \; {\pmb{\Upsilon}} = \lambda \, {\pmb{\Upsilon}} \, . \label{eval} \ee 

Like Boltzmann's collision operator, ${\mathcal L}$ is self-adjoint and, if we suppose that its spectrum is discrete (as it is for certain molecular interactions),  we have a countable, complete set of eigenfunctions.  It is central to the theory that
$\lambda =  0$ is a quintuple eigenvalue
of ${\mathcal L}$ corresponding to the five eigenfunctions $\psi^A$,  which span the null space (also known as center space) of 
${\mathcal L}$.  All the other eigenvalues are real and negative and their associated eigenfunctions span what we may call damped space (a.k.a. stable space).
\subsection{Weakly Nonlinear Theory} 
As suggested in subsection 5.1, the linear structure of the problem recalls center manifold theory (which has roots in kinetic theory) and so offers a way to think about the dimensional reduction in the passage from kinetic theory to fluid dynamics \cite{coul83}.        

In linear theory we may express ${g}$ as \be
{g}({\bf v}, t) = 
\sum_{A=0}^4 \alpha_A(t)\, {\psi}^A(\bf v)  + 
\sum_{B=0}^\infty \beta_B(t)\, {\varphi}^B(\bf v)  \label{exp}  \ee
where $ \alpha_A$ and $\beta_B$ are the expansion coefficients indexed for use in the invariant subspaces of linear theory\footnote{The eigenfunctions are the basis vectors in the phase space and may be regarded as fixed.} where $\psi^A$ and
$\varphi^B$ are the eigenfunctions of ${\mathcal L}$.  

The notation is becoming cumbersome so let us abbreviate
$\alpha_A$ and ${\psi}^A$ as ${\pmb{\alpha}}$ and ${\pmb{\psi}}$ and 
$\beta_B$ and $\varphi^B$ as ${\bf B}$ and ${\pmb{\varphi}}$. 
Then we may write \be
g = {\pmb \alpha}\cdot{\pmb{\psi}} +
{\pmb \beta}\cdot{\pmb{\varphi}}\, .
\ee

As in center manifold theory, we adopt a view based on the 
intuitive image that the fast (or damped) modes, ${\pmb{\varphi}}$,  come quickly into equilibrium with the surroundings while the slow modes, deformed by nonlinear coupling to the fast modes and to themselves, evolve slowly, taking the slavish fast modes along with them.  We could substitute the expansion for ${g}$ into (\ref{u}) and carry out suitable projections to get equations for the expansion coefficients, taking advantage of the orthogonality of the eigenfunctions of ${\mathcal L}$ \cite{foch70}.  But a more attractive approach is available when the problem has the features that are behind the marvel that is center manifold theory.   Whether explicitly or implicitly, this image is adopted in several subjects including chemical kinetics, nuclear reactions in stars and certain asymptotic methods as in \cite{dellar07}.  Each system constituted in the manner we have described is attracted to a slow subspace whose dimension is equal to the number of marginal modes of its linear theory. 
  
More precisely, we may say that the $\beta_B$ are functionals of the $\alpha_A$ in weakly nonlinear theory.  We call this the Bogoliubov ansatz while some know it as Stuart-Watson theory.  If we accept this outlook, then  
${\pmb{\beta}} = {\pmb{\mathcal B}}({\pmb{\alpha}})$, and $f$ 
becomes a function of ${\pmb{\alpha}}$ alone.   We can then turn the equation for $f$ into an equation for ${\pmb{\alpha}}$. 
I shall  not follow that path here since the procedure is spelled out in \cite{coul83}.  But an examination of the definition of 
${\pmb{\alpha}}$  will reveal that its components are in fact $\rho$, 
${\bf u}$ and $T$ up to constant coefficients that depend on the
normalization of the eigenfunctions.  Hence, the equations for the five components of ${\pmb{\alpha}}$ --- the coefficients of the expansion in $\psi^A$ ---  are the fluid equations.   The behavior of the fast modes is also derivable once the amplitudes of the slow modes are found.  

We discern an analogy to multiple
bifurcation theory here.  Indeed, one reason for this digression is to 
bring this point out.  If there are terms that tend to drive the system 
from equilibrium, that tendency will be resisted by coupling to the damped fast modes.  The departure from equilibrium is like
what would be called in nonlinear stability theory a bifurcation 
of codimension five.  And with these hints at how a system may
leave equilibrium, we rejoin the readers who skipped this section. 
\section{Closure}
\subsection{Preliminary remarks}
Our task now is to express ${\mathbb P}$ and ${\bf Q}$, defined
in ($\ref{presst}$) and ($\ref{heatf}$), in terms of 
$\rho$, $T$ and ${\bf u}$.  These latter three may depend on 
${\bf x}$ and $t$ which are effectively only parameters as far as 
${\mathcal C}$ is concerned.   We may obtain the desired expressions by seeking an approximate solution of the kinetic equation and using it to derive closure relations, as Hilbert proposed a century ago \cite{hilbert12}.  Since Hilbert's work, Boltzmann's equation has been the leading model in kinetic theory though several other models have been proposed for the collision term.  Being based on two-body collisions, Boltzmann's model is very specific and so has a crisp meaning.  The question of whether it suffices to consider only two-body collisions as the basic interaction of a particle with its surroundings has been intensively discussed and many believe that this is a good model at low density.

To see what the restriction to two-body interactions of the Boltzmann model entails let $\ell$ be the range of interaction of the particles and $\xi:=<|{\pmb{\xi}}|>$ be  their average peculiar
speed. Then the duration of a 
two-body interaction is of order $\ell/\xi$.  The chance that a third  body will intrude while a two-body interaction is in  process is measured by the nondimensional parameter 
$\ell/\left(\xi\varkappa\right)$.  When the medium is so rarefied that this parameter is very small, it may be safe to neglect three-body interactions.  

There are other issues that ought to be considered such as the
tendency of randomly moving particles to cluster.  And if the medium is too rarefied one may worry about the use of the continuum approximation.  But, as we have seen, continuum equations are a formal consequence of kinetic theory so the real question is whether we can find a closure that leads to adequate
results in such conditions.

To simplify the presentation of the derivation of a closure approximation, I shall use the relaxation model of the kinetic
equation rather than the Boltzmann model.  
The relaxation model, like most simplified collision models, is formulated with features of Boltzmann's model in mind.  Though I
shall not be concerned with the details of Boltzmann'a model, I shall adopt general conclusions that follow from it \cite{uhl63}.  And those inspire the treatment I give here.  But, when it comes to explicit calculations, I fall back on the relaxation
model of Boltzmann's equation --- it is very like the equation of radiative transfer that has been studied by astrophysicists since the early part of the last century.   It was only in 1954 that this form of transport equation was applied to material particles \cite{wel54, 
bhat54}. The use of the relaxation model allows us to avoid the cumbersome calculations that are needed to derive results from Boltzmann's equation without invoking the Chapman-Enskog iteration. 

\subsection{The relaxation model}
Hilbert sought approximate solutions of Boltzmann's equation by way of an expansion in the nondimensional number named for 
Knudsen, the Danish physicist who early studied rarefied gases.
We introduce a characteristic macroscopic time scale of the system, $\Theta$, and let the Knudsen number be
\begin{equation}
\varepsilon = \frac{1}{\varkappa\Theta}\, .
\end{equation}
But not even Hilbert succeeded in deriving anything more than the Euler equations with his expansion, and those lack dissipative terms.  Among other proposed approaches, those of Chapman and of Enskog have been most used \cite{uhl63, grad63}.   Also other collision operators have been been introduced, among the most successful being the relaxation model.  With this model, the work in calculating ${\mathbb P}$ and ${\bf Q}$ without using the C-E iteration becomes less onerous and its results seem comparable to those found with the Boltzmann model.

In keeping with with Boltzmann's results, it is natural to 
assume that the effect of collisions is to attract the system into a
state of local equilibrium given by the local Maxwell-Boltzmann distribution.   Then (\ref{kineq}) becomes   \begin{equation}
\varepsilon {\mathcal D}f = (f_0 - f).\label{relax}
\end{equation} 
Equation (\ref{relax}) causes $f$ to relax toward the $f_0$ chosen to be $f_M$ as specified in (\ref{f0}).
It is usual to treat $\varepsilon$ either as a
constant or as a function of $T$ in this model.  
\subsection{The Hilbert expansion}
We next expand $f$ in terms of $\varepsilon$ as  \begin{equation} 
f = \sum_{n=0}^\infty \varepsilon^n f_n\, . \label{HS}
\end{equation}
There are two parallel developments based on this expansion as reviewed by \cite{uhl63}: the development by Hilbert himself, which has been criticized \cite{caf83}, and 
the Chapman-Enskog method, aiming to 
expand the equations in power series instead of expanding the solution as Hilbert did \cite{grad63}. 
 
If we keep only the first term in the series, we obtain 
${\mathbb P} = {p {\mathbb I}}$ and ${\bf Q} = {\bf 0}$ where 
\be
p = \Re \rho T\ \label{EoS} \end{equation}
is the pressure and ${\mathbb I}$ is the unit dyad.
This approximation leads to the Euler equations for an inviscid fluid that conserves the specific entropy, defined as \begin{equation}
S = c_v \log\frac{p}{\rho^\gamma} \, , \end{equation}
where $c_v$ and $c_p$ are the specific heats and 
$\gamma = c_p/c_v\, .$

If we keep the first two terms in the series, we find, to order 
$\varepsilon^1$, that \begin{equation}
{\mathcal L}f_1 = {\mathcal D}f_0, \end{equation}
where ${\mathcal L}$ is the linearization of the collision operator
and ${\mathcal D}$ is the streaming operator defined as\begin{equation}
{\mathcal D} := \partial_t  +  v^i \partial_i  + a^j 
\frac{\partial}{\partial v^j}\, \end{equation}
where  ${v^i = \dot x^i}$ and ${a^i = {\dot v}^i}$. 
Then when we introduce the Maxwell-Boltzmann distribution we
find, on assuming that there are no external forces acting 
({\bf a}={\bf 0}),
\begin{equation}
{\mathcal D}f_0 = f_0{\mathcal D}\ln f_0 = - f_0 \left[{\mathcal D} \ln \rho +
\left(\frac{{\bf c}^2}{2RT} - \tfrac{3}{2} \right) {\mathcal D}\ln T
+ \frac{1}{RT} {\bf c}\cdot {\mathcal D}{\bf u}\right] \ . \label{f1}
\end{equation}
The advantage of working with the relaxation model is that instead of having to invert the linearized collision operator of Boltzmann we have immediately the solution \be
f_1 = -{\mathcal D}f_0.\label{f_1}
\ee
\section{${\mathbb P}$ and ${\bf Q}$}
In the Chapman-Enskog approximation one replaces the terms in ${\mathcal D}$ that involve $\partial_t$ with the corresponding expressions from the Euler equations.  However, we do not iterate in this way since the Hilbert expansion is asymptotic and not convergent \cite{grad63}.  Instead we keep all the terms in (\ref{f1}) and by straightforward integrations find
\begin{equation}
{\mathbb{P}} = p \mathbb{I}
- \left[\mu\left({\frac{D \ln T}{Dt}} + \tfrac{2}{3}
\nabla \cdot {\bf u} \right)\right] \mathbb{I} - \mu \, \mathbb{E}
        + \mathcal{O}(\epsilon^2),
        \label{P}
\end{equation}
  
\begin{equation}
E_{ij} = {\frac{\partial u_i}{\partial x_j}} + {\frac{\partial u_j}
        {\partial x_i}} - \tfrac{2}{3} \nabla \cdot {\bf u} \, \delta_{ij}
        \label{E}
\end{equation}
\begin{equation}
{\mathbf{Q}}= -\eta\nabla T  - \left[\tfrac{5}{2}\mu
 {\frac{D{\bf u}}{Dt}} + \eta T \nabla\ln p\right] + 
 \mathcal{O}(\epsilon^2)
        \label{Q} \end{equation}
where $\mu=\epsilon p$ and $\eta=\tfrac{5}{2}\mu \Re$.  The terms in square brackets in
(\ref{P}) and (\ref{Q}) are suppressed when the Chapman-Enskog
approximation is used.  Those terms are easily seen to be 
${\mathcal O}(\varepsilon^2)$ like the higher order terms in
the expansion.  The problem is that those C-E expressions for 
${\mathbb P}$ and ${\bf Q}$ are often used even when 
$\epsilon$ is not very small.  The new terms then make a significant difference.  

The decision as to which of the asymptotically equivalent versions of the fluid equations should be used
is best made by comparison with experiment at large Knudsen number.
Such comparison for the thicknesses of shock waves for moderate Mach numbers favors the present results \cite{preIII}. 
Also, our results for the 
phase speed of ultrasound waves are in reasonably good agreement with experiment while those from the N-S equations fare poorly in that comparison when the Knudsen number exceeds unity \cite{preII}.  (Differences between those results are mainly related
to the different values of the Prandtl number obtained with the two
models)

Neither the N-S results nor our present results at first order agree with measurements of  the damping length of sound waves when the Knudsen number is greater than unity.   Simply going to the next order does not solve that problem \cite{spi03}.  What seems to work is to go to next order and to resum the relevant terms.  The problem in doing this is that, already in ${\mathcal O}(\varepsilon^2)$, there is a plethora of
terms and it is not yet clear which must be retained.  That matter
is too complicated for discussion here \cite{unn66, cherad, rose89, slem97} so I will briefly indicate an alternate related approach. 
 
From the forgoing formulae (with no external forces) we find that \be
\dot S := {\mathcal D}S = C_v\left({\mathcal D}\ln T + \tfrac{2}{3} \nabla
\cdot {\bf u}\right)
\ee
where ${\mathcal D} := \partial_t  + {\bf u}^i \partial_i \, .$
From this and (\ref{P}) we obtain \be
{\mathbb P} =
 \left(1 - \frac{\varepsilon}{C_v} \dot S\right)p{\mathbb I} 
 - \mu {\mathbb E} + {\mathcal O}(\varepsilon^2).  \ee
But ${\dot S}/C_v = \dot p/p - \tfrac{5}{3}\dot \rho/\rho$
and so the trace of ${\mathbb P}$ is $3p(t-\varepsilon) +5\mu \nabla \cdot {\bf u}$ where $p(t-\varepsilon)$ means $p$
evaluated at time $t-\varepsilon$.  And so we are dealing with
what is called a delay equation \cite{erneux}.  These can produce
damping, either negative and positive, depending on the parameter values.  For a proper treatment of this issue it will be necessary to
go to higher order in $\varepsilon$ with this approach before
comparison with experiment becomes sensible.  But for here and
now space and time are running out and I must postpone that elaboration to another time and place.  Anyway, the deeper issue is that of causality and, for that, we now turn to the more engaging relativistic case.\footnote{As one of my favorite teachers, G.E. Uhlenbeck, often exclaimed ``the description must be causal, it must fulfill the causal property.  If it doesn't have that then it is not worth a damn.''}

 \bigskip  

\section*{\centerline{\bf II. Causality and Cosmology}}

\section{The Transfer Equation}
Again, we work with a model gas with a large
number of identical, structureless particles but, this time, we
characterize them by their momenta rather than their velocities.   We work in a phase space but, this time, we are in four-dimensional spacetime and in momentum space.  This choice allows us to deal with photons as readily as with material particles.  (But we leave 
spin out of account.)  The transport equation that we adopt here has been used for both kinds of particle since L.H. Thomas used it in his study of photon transport in 1930 \cite{tho30, lind66, stew71, and72}. 
  
With $x^\mu$ as the coordinates in spacetime, with Greek indices running through $0,1,2,3$ and with  $p^\mu$ as the 
four-momentum, the equation governing the evolution of
the phase-space particle density, $f(x^\mu, p^\nu)$, is \cite{tho30}
\be
 p^\mu f_{,\mu}=\alpha - \kappa f \, .
\label{1} \ee 
Here particles are scattered into the volume element $d p^\mu$ containing $p^\mu$ at the rate $\alpha$ by interactions with the ambient medium.  And particles are scattered out of that volume element at a rate $\kappa f$.  According to preference, $f$ it may be a scalar (as here) or a scalar density.   The speed of light and Planck's constant are unity.  For the time being, we'll be living in Minkowski space but the trip into curved space that we'll make later on will be an easy one requiring us only to replace commas by semicolons. (For a more geometric view of the physics see
\cite{lind66}, for example.)  

As in the nonrelativistic case, the local equilibrium solution is \be 
f= f_{0} = \frac{\alpha}{\kappa} \, . \label{KP1} \ee
(Here, the subscript zero is of course not an index.) 
And the transformation rule for $\kappa$ given by Thomas is \be
\kappa = \hat\kappa u^\mu p_\mu \ee
where $\hat \kappa$ is the value of $\kappa$ in the local rest frame.  With $\epsilon = 1/\hat \kappa$ (\ref{1}) then takes the relaxation form \be
\epsilon p^\mu f_{,\mu}=u_\nu\,p^\nu(f_{0} -  f)\, . \label{relax1} \ee Following in the wake of Thomas \cite{tho30} this equation has been used in the study of photon transport \cite{and72} and adopted also for studying transport in material particles \cite{and74a, krem, cqg}.  The merit of the relaxation model is that it offers simplicity when, as here, $\hat \kappa$ is assumed independent of $p^\mu$ and it lacks the restriction of Boltzmann's model to binary collisions.   

When the particle number is conserved, we find from (\ref{1}) that  \be
\int p^\mu f_{,\mu}dP=0. \ee
Since $p^\mu$ is independent of $x^\mu$, this leads to \be 
N^\mu_{\ ,\mu}=0 \label{7} \ee 
where   
\be N^\mu=\int p^\mu fdP\; \ .
\label{8} \ee 
Here $dP$ is the invariant volume element in momentum space \cite{lind66, ber88}, 
\be dP=d^3p/e \label{dP}\ee                                                                                            where $d^3p$ is the three-dimensional volume element in momentum space and $e$ is the
particle energy.
 
Then, when we multiply (\ref{1}) by $p^\nu$ and integrate, we find \be
T^{\mu\nu}_{\ \ , \mu}=0 \label{11} \ee 
where \be T^{\mu\nu}=\int p^\mu p^\nu f dP \ .
\label{12} \ee 
\section{In component form}
It is convenient to decompose the basic quantities into component form with the help of the projection operator \be
h^{\mu \nu} = g^{\mu \nu} - u^\mu u^\nu \ee
where $ g^{\mu \nu}$ is the metric tensor (as yet, the Minkowski metric).  The component of $N^\mu$ along $u^\mu$ may be identified as the number density of particles
in the rest frame, \be
N = u_\mu N^\mu \ ,  \ee  
and the number density current is \be 
J^\mu = h^{\mu \nu} N_\nu \ . \ee

We further introduce the decomposition 
$p^\mu = (e, {\bf p})$
where  ${\bf p}$ is the three-momentum and $p=|{\bf p}|$.  Then, we let $\hat e = u_\mu p^\mu$
be the particle energy in the local rest frame of the fluid.  We 
also define $\hat p$ such that \be
\hat p^2 = h^{\mu\nu}\, p_\mu p_\nu = m^2 - \hat e^2 \, . \ee
Next we introduce $l^\mu$ such that \be
l^\mu = \frac{h^{\rho\mu}\, p_\rho}{\hat p} \qquad {\rm and}
\qquad
p^\mu = \hat e u^\mu + \hat p\, l^\mu \, . \label{split} \ee
We see then that $l^\mu l_\mu  = -1 $
and $l^\mu u_\mu = 0$.
  
The stress tensor may then be expressed as  \be
T^{\mu \nu} = E u^\mu u^\nu +  F^\mu u^\nu + F^\nu u^\mu + 
P^{\mu \nu}   \  .  \label{prtens} \ee
where  \be
E = \int \hat e^2fdP \qquad
F^\mu =  \int \hat e \hat p l^\mu f dP   \qquad P^{\mu \nu}= \int \hat p^2 l^\mu l^\nu fdP
 \ . \label{comp} \end{equation}
Also   \be
E= u^\mu u^\nu T_{\mu \nu} \qquad
F^\mu =  h^{\mu \nu} u^\rho T_{\nu \rho} \qquad P^{\mu \nu}= h^{\mu \rho} h^{\nu \sigma} T_{\rho \sigma}
\ . \label{decomps}
\end{equation}
Then, (\ref{7}) and (\ref{11}) become
\ba & & u^\mu N_{,\mu}+N\vartheta+J^\mu_{\ ,\mu} = 0 \label{cont} \\ & &
u^\mu(Eu^\nu)_{,\mu} + E u^\nu\vartheta +(F^\mu u^\nu+F^\nu
u^\mu)_{,\mu}+P^{\mu\nu}_{\ \ ,\mu}=0 \label{mo} \ea 
where $\vartheta=u^\mu_{\ ,\mu}$.
\section{The Landau-Lifshitz frame}
It remains to choose a reference frame by giving five relations among the five macroscopic fields.     
We adopt the five conditions that Landau and Lifshitz impose to fix the frame
choice, $u^\mu$.  These are                      
\be u_\mu N^\mu=u_\mu N_{0}^\mu
\label{19}
\ee and \be u_\mu T^{\mu\nu}=u_\mu T_{0}^{\mu\nu}
\label{20}
\ee where $N_{0}^\mu$ and $T_{0}^{\mu\nu}$ are the appropriate moments of the local equilibrium distribution $f_{0}$. 

With isotropic $f_{0}$  the number current and the energy flux both vanish in local equilibrium.  Equations (\ref{19}) and (\ref{20}) then lead to 
\be N \equiv  N_{0}; \qquad E  \equiv E_{0}; \qquad
F^\mu \equiv  0\, .  \ee 
Thus, in going to the Landau-Lifshitz frame we transform away the energy flux and reduce (\ref{mo}) to 
\be u^\mu(Eu^\nu)_{,\mu} + E u^\nu\vartheta +P^{\mu\nu}_{\ \ ,\mu}=0\, , \label{LL2} \ee 
which may be written as \be 
Eu^\mu \vartheta+(Eu^\mu \dot)+P^{\mu\nu}_{\ \ ,\nu} =0
\label{27}\ee
where $(\ \dot) := u^\mu(\ )_{, \mu}$
can be broken down into components.
On projecting (\ref{27}) in the direction of $u_\mu$, we get the
energy conservation equation \be 
\dot E + E\vartheta +  u_\mu P^{\mu\nu}_{\ \ ,\nu}=0\; .
\label{28}
\ee 
When we project this with $h_{\rho\mu}$ we obtain the equation of motion
\be h_{\rho\mu}\left(E\dot u^\mu+P^{\mu\nu}_{\ \  ,\nu}\right)=0\; .
\label{29} \ee      
\section{Approximating $f$}
\subsection{The ultrarelativistic gas}
We now derive a closure relation without invoking the C-E iteration  
for a gas consisting of one type of particle with no
internal degrees of freedom.    To simplify the presentation even
further, we presume that the particles are ultrarelativistic so
that their masses may be left out of account in a description that
might be appropriate for a  photon-dominated medium.   This approach leads to a closure approximation that is representative of more general situations without requiring the arduous calculations that situations with complicated particle mixes would call for.  And even in the case of the photon gas, interactions are possible since
energetic photons may scatter off one another
by creating virtual $e^+\negthinspace\negthinspace-\negthinspace e^-$ pairs.  
 
The commonality of this example with studies of radiative transfer
in material media is helpful in other ways. Just as for photons, we do not assume that 
the particle number is conserved (though we do not 
explicitly include quantum mechanics).  Hence 
we consider simply 
\be T^{\mu\nu}_{\ \ ,\mu}=0\, , \ee
to which our fluid dynamical equations boil down.   
Then, we complement equations (\ref{28}) and (\ref{29}) with an expression for $P^{\mu\nu}$
in terms of $E$ and $\dot E$.   

To find a closure relation for $P^{\mu\nu}$ we
introduce a series expansion of $f$ in terms of $\epsilon$ into 
(\ref{relax1}).  Then we use that to evaluate and relate the relevant moments of $f$.  For relativistic radiative fluid dynamics, we have carried this 
procedure out in~\cite{cherad}.   There we took $u^\mu$ as
the velocity of the ambient medium and treated it as known.
In the present example, there is no background medium and we 
are working with a gas of ultrarelativistic particles whose masses 
we neglect as for photons. 
We now have in mind a $u^\mu$ that is the appropriate velocity field of the fluid itself even in the case of a photon gas.  We then obtain a pressure tensor of the same form as given in \cite{cherad}.
 
A nice simplification of the ultrarelativistic problem is that a
particle's energy is equal to the magnitude
of its three-momentum.  Then, in thinking about the macroscopic aspects, it is useful to
introduce the null vector $n^\mu$ such that $n^\mu \hat e = p^\mu$ and, as in (\ref{split}), 
to decompose it into \be
n^\mu = u^\mu + l^\mu .\ee
Thus,  as in (\ref{prtens}), the components of the stress tensor may be expressed in terms of the quantities defined in (\ref{comp}). 
\subsection{The expansion of $f$}
To develop an expression for $P^{\mu \nu}$ by seeking an approximation for $f$, we return to the relaxation model 
(\ref{relax1}) with $p_\mu p^\mu = 0$.
We then seek an approximate solution in the form \be
f = f_{0} + \epsilon f_1 + {\cal O}(\epsilon^2)  \label{exp} \ee
where the naught is again a subscript not an index and we see that \be
f_1 = - n^\mu f_{0,\mu} . \ee
To evaluate the stress tensor we must now choose an $f_{0}$.
For the present example we adopt the equilibrium distribution 
for a gas of bosons with zero chemical potential:
\be
f_{0}=[e^{\beta u^\mu p_\mu}-1]^{-1}\, .
\label{30}
\ee
We use this expression in a local sense and allow that 
both $\beta$ ($=1/T$ in suitable units) 
and $u^\mu$ may depend weakly (to use this dangerous term) on $x^\mu$.   

The relaxation equation is always attracting $f$ toward the evolving equilibrium whose changes are connected to
$f$ itself through the matching condition,  
\be \int fdP=\int f_0dP\, .\ee
 
To find the variation of $f_{0}$ with $x^\mu$ we use
 the chain rule to write \be 
\partial_\mu f_{0} = \left(T_{,\mu} \partial_T + \hat e_{,\mu} \partial_{\hat e}\right) f_{0}\, . \label{chain} \ee
In higher orders, things become more complicated
(as in the nonrelativistic case 
\cite{spi03}).   For nonzero rest mass, we 
would also need to include a chemical potential.  

A further  simplification is that, as for the photon gas, $f_{0}$ depends on $\hat e$ and $T$ only through the ratio
$\hat e/T$  (Stefan's law) so that we may write \be
\partial_\mu f_{0}= \left(T_{,\mu} - T l^\rho u_{\rho,\mu}\right) \partial_T f_{0} \ . \ee
And so we conclude that, to first order accuracy in $\epsilon$, \be
f = f_{0} - \epsilon n^\mu  \left(T_{,\mu} - T l^\rho u_{\rho,\mu}\right) \partial_T f_{0} + {\cal O}(\epsilon^2) \; .
\label{1storder} \ee
\section{The Closure Approximation}
\subsection{Zeroth order}
In leading order, with $f=f_{0}$, we may use the last of  
(\ref{comp}) to evaluate the pressure tensor, which becomes  \be
P^{\mu \nu}  = \int \hat p^2l^\mu l^\nu f_{0} dP .
\ee
Because the medium is locally isotropic in zeroth order, the angular integral in momentum space involves only products of the $l^\mu$.  Such integrals are 
evaluated by expressing the results in terms of $u^\mu$ and
$g^{\mu\nu}$ and making suitable choices for the coefficients.  We find that \be
\int l^\mu l^\nu d\Omega = -  \frac {4\pi}{3} h^{\mu \nu} \ .\ee
We then obtain (in zeroth order) the closure relation \be
P^{\mu \nu}  =   - Ph^{\mu\nu} \ee  
where  with (\ref{dP}) we obtain \be
P =  \frac{4\pi}{3} \int \frac{\hat p^4}{\hat e} f_{0} d\hat p \ee 
which is identified as the pressure.
On using the first of  (\ref{comp}), we see that \be
P = \frac{1}{3}E\ , \label{state} \ee
which is the equation of state for equilibrium radiation,
where $E$ is the energy density.  Moreover, for the equilibrium 
(\ref{30}) we know
that $E=aT^4$, where $a$ is the radiation constant, so that serves to tie in the temperature. 
We also obtain the equations of motion for a perfect fluid 
at this order but we have no need of them here.
\subsubsection{First order}
Next we introduce (\ref{1storder}) into (\ref{comp}).  On performing the angle integral we find that \be
\int l^\mu l^\nu l^\rho l^\sigma d\Omega = \frac{4\pi}{15}\left( h^{\mu \nu} h^{\rho \sigma} 
+ h^{\mu \rho} h^{\nu \sigma} + h^{\mu \sigma} h^{\nu \rho}\right)\, . \ee
So we obtain on using the various definitions that \be
P^{\mu\nu}=- P h^{\mu\nu} + 
h^{\mu\nu}\mu \left(\frac{\dot E}{E}+
\frac{4}{3}\vartheta\right)+\Xi^{\mu\nu}   \label{30.5}  \ee 
in which \be
\Xi^{\mu\nu}=\frac{4\mu}{5}\tau^{\mu\nu\rho\sigma}u_{\rho,\sigma}
\label{31}
\ee 
is the viscous shear stress tensor where  \begin{equation}
\tau^{\mu\nu\rho\sigma}=h^{\mu\rho}h^{\nu\sigma}+
h^{\mu\sigma}h^{\nu\rho}-\frac{2}{3}h^{\mu\nu}h^{\rho\sigma} 
\label{tau} \end{equation}
and $\mu = P\epsilon$ is  the viscosity.

The middle term on the right side of (\ref{30.5}) does not appear in
the C-E results, which apply only for very short mean flight times.  This term is central to the difference of our approximation from the usual first-order results; it is a process-dependent term that allows for deviations from equilibrium.

To recover the first-order closure approximation found by using the
Chapman-Enskog procedure, we need only neglect our extra term,
which is of order higher than the first since $\mu$ is proportional to $\epsilon$.  But that truncation leads to acausal equations as noted by Israel \cite{isr63, isr76} who (as did others) preferred to go to second order to regain causality.  But among the regained terms in second order are those discarded in the C-E iteration.
 
Other approaches to deal with the causality failure in the heat equation exist both classically \cite{jbk04} and relativistically 
\cite{car}.  Yet the key question is, how good are the results quantitatively in the various approaches?   For answer to this question, we have elsewhere turned to the nonrelativistic case for which empirical tests are available.   These favor the avoidance of the C-E as reported in \cite{spi03,che00}.  
\subsection{The Fluid Equations}
To complete our statement of the basic equations, we note that
the projections of $P^{\mu\nu}_{\ \ \,,\nu}$ on $u_\mu$ and $h_{\rho\mu}$ are 
\be u_\mu P^{\mu\nu}_{\ \ ,\nu}=\frac{1}{3}\vartheta(E-\epsilon Q)+\Xi^{\mu\nu}_{\ \ ,\nu} u_\mu
\label{32}
\ee  and \be 
h_{\rho\mu}P^{\mu\nu}_{\ \ ,\nu}=\frac{\dot
u^\mu}{3}h_{\rho\mu}(E-\epsilon
Q)-\frac{1}{3}\delta_{\ \rho}^\mu(E-\epsilon
Q)_{,\mu}+h_{\rho\mu}\Xi^{\mu\nu}_{\ \ ,\nu} 
\label{34} \ee 
where 
\be Q=\dot E+\frac{4}{3}E\vartheta \ .\label{33} \ee

On substituting (\ref{32}) and (\ref{34}) into (\ref{28}) and (\ref{29}) respectively, we find \be
Q(1-\frac{1}{3}\vartheta \epsilon)=-u_\mu\Xi^{\mu\nu}_{\ \ ,\nu}
\label{35} \ee 
and, since $h_{\mu \sigma}\dot u^\sigma=\dot u_\mu$,  \be 
\dot u_\mu(4E-\epsilon Q)
=(E-\epsilon Q)_{,\mu}-3h_{\rho\mu}\Xi^{\rho\sigma}_{\ \ ,\sigma} .\label{36} \ee
  
Since (\ref{30.5}) involves a derivative of $E$ as well as $E$ itself, 
it does not have the usual look of a closure relation ---
it is process-dependent.   We may replace $\dot E$ in (\ref{30.5}) by using (\ref{28}).  That would
bring in $u_\mu P^{\mu \nu}_{\ \ ,\nu}\; ,$ which we could replace using (\ref{32}) with (\ref{33}).  That in its turn brings back $\dot E$ but this time as a term ${\mathcal O}(\epsilon^2)$.  If repeated indefinitely, this procedure brings in terms of all orders in powers of 
$\epsilon$.  That is the reason that one may expect that (\ref{30.5}) 
can be a significant improvement over the standard closure formulae.   However, we prefer to leave well enough alone here and to retain the closed form with $\dot E$ rather than the infinite sum based on $E$ alone.  
This sequence of substitutions is different than the 
C-E procedure as there is here no truncation; in any case, we do not use it but merely mention it for background.
 
\section{The Einstein Equation}
We next combine (\ref{35}) and (\ref{36}) (derived by eschewing the C-E iteration) with the Einstein field equation, 
\be G^{\mu\nu}= T^{\mu\nu}\, ,\label{37} 
\ee 
under the conditions of homogeneity and isotropy.  We have 
a flow described by five equations for the five unknowns in $E$, $H (=\vartheta/3)$ and $u^\mu$ and ask
what modification the equations found here may introduce into the study of a self-gravitating medium in the case where the mean flight time of the constituent particles need not be infinitesimal and 
noneqiulibrium may be expected to prevail    

In an isotropic medium $\Xi^{\mu\nu}\equiv 0$ and, with 
$H=\dot R/R$, (\ref{35}) and (\ref{36}), become 
\be Q(1-H\epsilon)=0
\label{381}
\ee 
\be  \dot u_\mu(4E-\epsilon Q)= (E-\epsilon Q)_{,\mu}\ .
\label{39}
\ee 
 
In the currently preferred cosmological model with 
zero curvature, the Einstein equation (\ref{37}) where $E$ is the energy density becomes \cite{wei72}
\be  E=3H^2\; 
\label{41}
\ee 
With the rest-mass density 
left out of account in this case of a medium dominated by ultrarelativistic particles (\ref{381}), (\ref{39}) and (\ref{41})
admit the following two solutions: 

\begin{enumerate}

\item One solution to (\ref{381}) is $Q=0$.   
In that case (see (\ref{33})), we have
\be \dot E+4E H=0\, ,
\label{42}
\ee 
which gives the familiar result that $ER^4$ is  constant.
If we assume that, as in 
the equilibrium of a photon gas, 
$E\propto T^4$,
we have $T\propto R^{-1}$, the well-known cooling law for equilibrium radiation in an expanding medium.   With $Q=0$, we conclude from (\ref{39}) that $u_\mu \propto R^{-1}$ and we have recovered some key features of the Friedmann solution.
 
\item The other possible solution of (\ref{381}) is 
\be H\epsilon = 1. \label{Kn} \ee 
This is really an approximate solution that holds up to errors
of order $\epsilon^2$. The nondimensional parameter, 
$\epsilon H$, is the ratio of the mean free flight time of the particles to the time scale of the macroscopic expansion and is a local Knudsen number in fluid dynamical terminology.
 
Though it is not excluded that we may also have $Q=0$ in this
situation, we omit that case for brevity.  The temperature is then not forced to behave in the way that it does in local equilibrium in the FRW model and it is undetermined as yet.
Though we have assumed that $\epsilon$  does not depend on 
$p^\mu$ explicitly, it may depend on local (in time) macroscopic properties of the medium such as temperature and pressure.  Those variables, in their turns, will generally vary with $R$.   Once all those dependences are specified, equation (\ref{Kn}) becomes a differential equation for $R$.  A formal solution may be written
for $R(t)$  but nothing explicit can be said until the dependence of $\epsilon$ on local physics is specified.  In the somewhat unrealistic conditions of constant $\epsilon$, $H$ would also be (nearly) constant.   The solution
\be R\propto \exp(Ht), \label{infl}\ee 
with $t_0=-\infty$ and $R_0=0$ does have a certain simplicity
but its message that the e-folding time of $R$ in solution 2 is the mean flight time of the particles will require some further thought.    
\end{enumerate}

These possibilities cannot be properly evaluated without a more explicit particle model.  The main point for now is that the truncation introduced in the Chapman-Enskog procedure has filtered out one of the two solutions of our equations for the 
expansion of a relativistic, self-gravitating gas.   And that second
solution has a different physical origin than the usual ones.

Solution 1 may be considered geometrical in origin; after all, expansion can occur even in a cosmology without matter.  
Solution 2 however is matter-driven as we see from the close connection between the expansion rate and the mean flight time.  
In that sense, solution 2 is analogous to a shock wave whose geometric thickness is typically within an order of magnitude of the mean free path of the particles in the fluid \cite{lie47}.   But, in solution 2, this shock-like behavior takes place in time rather than in space so that the expansion time scale is of the order of the mean flight time of the constituent particles.  In the two situations, the macroscopic inhomogeneity (whether spatial or temporal) is 
strongly influenced by the microscopic behavior of the particles. \subsection{Entropy Production} 
In solution 1, once the medium is in local thermodynamic equilibrium, the expansion will not destroy the equilibrium if the particles are ultrarelativistic.  However, since the expansion rate
and the mean flight time are comparable in solution 2, we cannot
expect thermodynamic equilibrium to be established in that case.  Yet the relatively infrequent collisions that do occur will generate entropy.  The only macroscopic effect in the present model that 
can account for this is the changing volume that, in this case, is a disequilibrating effect occurring through the volume (or bulk) viscosity which may generate entropy.   

For a
rough estimate of the rate of entropy production in solution 2 
we treat $\epsilon$ as constant.  In that case, $H$ is roughly constant and the expansion is exponential as in (\ref{infl}). 
For an estimate of the rate of entropy production we use the formula that has
been computed in the study of radiative fluid dynamics in \cite{cherad},  \be 
 \dot {\cal S}=\xi\vartheta^2\ee
 \be \xi=\frac{4\epsilon E} {T}\left(\frac{1}{3}+
\frac{\dot T}{T\vartheta}\right)^2\ee
where ${\cal S}$ is the entropy density.   
This is a single temperature result that
resembles that found by Weinberg \cite{wei72} for a two-temperature medium.
 
Since $\vartheta=3H$, the rate of entropy generation 
(which vanishes in solution 1) is 
 \be \dot{\mathcal S}=\frac{4\epsilon E}{T}\left(\frac{\dot R}{R}+\frac{\dot T}{T}\right)^2,
\label{44} \ee 
For this illustration, let us imagine that $\epsilon H$ passes through
unity in the early stages of expansion so solution 2 becomes viable
for a time.  If that happens, there could occur a transition period in which  the macroscopic evolution is driven by the interplay of the particle interactions and the large-scale expansion.  
It is internally consistent to take $\epsilon,\ T$ and $H$ as approximately constant before an exchange of stabilities causes
solution 2 to give way to solution 1, for instance.    During the time interval when $\epsilon H$ is of the order of unity, solution 2 would then lead to an  entropy generation rate of the order of \be
\dot {\cal S} = \frac{4EH}{T}.\ee
In that case, we estimate the entropy at time $t_1$ to be 
 \be {\cal S}= (t_1-t_0) \dot {\cal S}  \; ,
\label{45} \ee 
if solution 2 applies from $t_0$ to $t_1$. 
To find the total comoving entropy we multiply by $R^3$ 
\cite{maa95}.  Then, at the hypothetical instant at
which we presume that solution 2 may return the baton to solution 1, we find that (see (\ref{33}) 
  \be 
R^3{\cal S}=R_0^3e^{3H(t_1-t_0)}Q t_1\, .
\label{46} \ee 

To get some idea of what kind of quantitative effect to expect we 
compute the entropy generated during the conventional estimate
of the time interval sometimes adopted for allowing inflation to cure some cosmological ills.  Thus we take the previously suggested 
\cite{kol90} $t_0=10^{-35}$s and ask what the entropy is at
$t_1=10^{-32}$s.  For this, we adopt for $H$ and $T_0$ the values  $6\times 10^{33}$s$^{-1}$ and $10^{28}$K, respectively.  We also have $R_0=ct_0$.  The total comoving entropy generated in the interval $t_0$ to $t_1$ is then of the order 
$10^{88} J/K$.   This estimate is comparable to current 
estimates of the value of the entropy of our universe~\cite{ber88} and has also been 
found when the entropy generation is ascribed to matter/antimatter annihilation \cite{ber88} or to bulk viscosity \cite{maa95} in the standard solution.  Other times and parameter values could have been chosen; we have put values found in the literature into our estimates for comparison.   If indeed solution 2 did arise in those very early times it may have contributed to the heating of the universe.

{\bf Acknowledgments.} In preparing these notes I have
benefited from helpful comments of several people among whom are Steve Childress, Joe Keller, Sara Solla, Charles Tresser,
and Phil Yecko.  I am grateful for Milena Cuellar's aid with 
the style file and to Lenny Smith and Phil Yecko for their patient help with technical aspects of manuscript production.

\end{document}